# Shear driven formation of nano-diamonds at sub-gigapascals and 300 K

Yang Gao[a], Yanzhang Ma[a,*], Qi An[b], Valery Levitas[c,d], Yanyan Zhang[a], Biao Feng[c,e], Jharna Chaudhuri[a], and William A. Goddard III[f]

[a] Department of Mechanical Engineering, Texas Tech University, 2500 Broadway St. Lubbock, Texas 79409, USA.

[b] Department of Chemical and Materials Engineering, University of Nevada, 1664 N Virginia St, Reno, NV 89557, USA.

[c] Department of Aerospace Engineering, Iowa State University, Ames, IA 50011, USA

[d] Division of Materials Science and Engineering, Ames Laboratory, 311 Iowa State University, Ames, IA 50011, USA.

[e] Los Alamos National Laboratory, Los Alamos, NM 87545, USA

[f] Materials and Process Simulation Center, California Institute of Technology, 1200 E California Blvd, Pasadena, California 91125, USA.

**Corresponding Author:** Yanzhang Ma

Department of Mechanical Engineering, Texas Tech University

Lubbock, TX 79409

E-mail: y.ma@ttu.edu

Phone: (806)834-3633

**Abstract** The transformation pathways of carbon at high pressures are of broad interest for synthesis of novel materials and for revealing the Earth's geological history. We have applied large plastic shear on graphite in rotational anvils to form hexagonal and nanocrystalline cubic diamond at extremely low pressures of 0.4 and 0.7 GPa, which are 50 and 100 times lower than the transformation pressures under hydrostatic compression and well below the phase equilibrium. Large shearing accompanied with pressure elevation to 3 GPa also leads to formation of a new orthorhombic diamond phase. Our results demonstrate new mechanisms and new means for plastic shear-controlled material synthesis at drastically reduced pressures, enabling new technologies for material synthesis. The results indicate that the micro-diamonds found in the low pressure-temperature crust could have formed during a large shear producing event, such as tectonic rifting and continued plate collision, without the need to postulate subduction to the mantle.

High pressure has been proved essential in the formation of diamond (D) since its first synthesis at 10 GPa and 1000-2000 K.[1] The recent transformation to transparent and much harder nano-polycrystalline D from direct conversion of graphite (G)[2] and twinned D from onion carbon[3] requires even higher pressures (12 GPa and 2000 ºC). Application of a catalyst can reduce the transformation pressure but only to about 5 GPa at 1500 K.[4, 5] Recent experiments showed that reduced distance between carbon atom under high pressures, leads to formation of $sp^3$ σ-bonds, but after unloading the bonding returns to $sp^2$ π-bond character.[6] The simultaneous application of shear with pressure dramatically changes the transformation pathway, e.g., from a displacive to a reconstructive transformation of hexagonal to wurtzitic boron nitride.[7]

Here, we demonstrate that application of shear on G under low pressure at 300 K can transform G to several D forms and to other phases, thereby revealing new pathways for phase transformation while retaining the high-pressure phases after releasing the pressure. The transformation pressures to hexagonal D (hD) and cubic D (cD), determined with *in-situ* synchrotron X-ray, are respectively 0.4 GPa and 0.7 GPa, which are 50 to 100 times lower than under quasi-hydrostatic conditions.[8, 9] After additional pressure-shear processing we show that the quenched samples contain cD, orthorhombic D (ortho-D), fullerenes, fragmented G networks, and amorphous phases. We also report theoretical studies at multiple scales to reveal the mechanisms for this drastic reduction in transformation pressure and formation of diverse phases. Our results suggest a new mechanism of micro D formation in the Earth at the lower pressure-temperature crust conditions without the need to postulate subduction to the mantle. Instead, large shear produced during the historical tectonic activities could be responsible for formation of micro D.

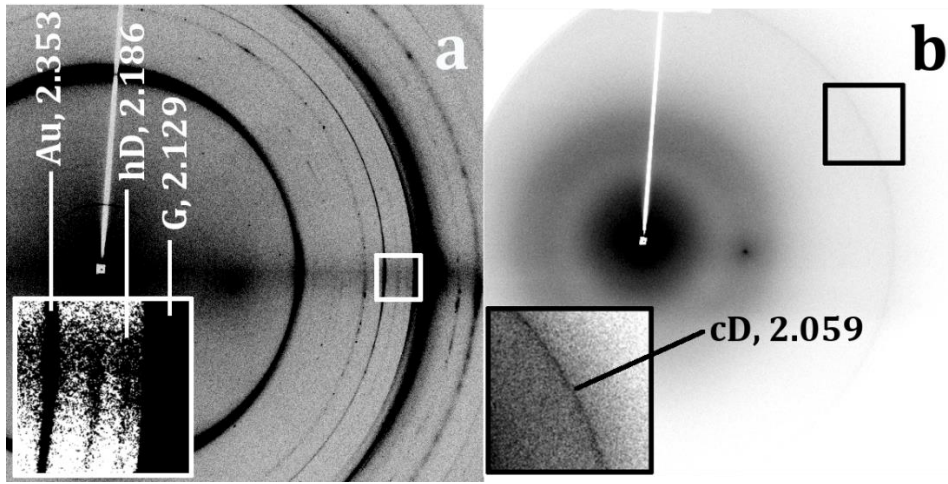

*Fig. 1: The X-ray diffraction image of compressed and sheared graphite. a, in-situ at 0.4 GPa after 45˚of anvil rotation; b, quenched to room pressure. Au, gold, hD, hexagonal diamond, cD, cubic diamond, G, graphite. The inset images are respectively the enlargement to the marked rectangular area.*

*Formation of D phases.* In our experiments, we first applied an axial load to a G sample in the rotational anvil apparatus (see Supplementary Materials (SM)) to attain a pressure less than 0.1 GPa followed by anvil rotation operations to generate plastic shear strains; then we increased both the load and the shear repeatedly before finally releasing the load to quench the sample to room pressure. We characterized the sample using X-ray diffraction after each operation, and analyzed after quenching. In the first few shear operations, we observed a gradual splitting of the peak corresponding to the G (002) planes as a result of the shear introduced pressure elevation. The (002) peak of G then started to disappear gradually, at which stage we observed the appearance of diffraction spots at 2.186 and 2.059 Å (Fig. 1a), which indicates the formation of hD. At pressure of 0.7 GPa and after quenching, we observed only the 2.06 Å peak, which shows that a cD crystal has been formed (Fig. 1b).

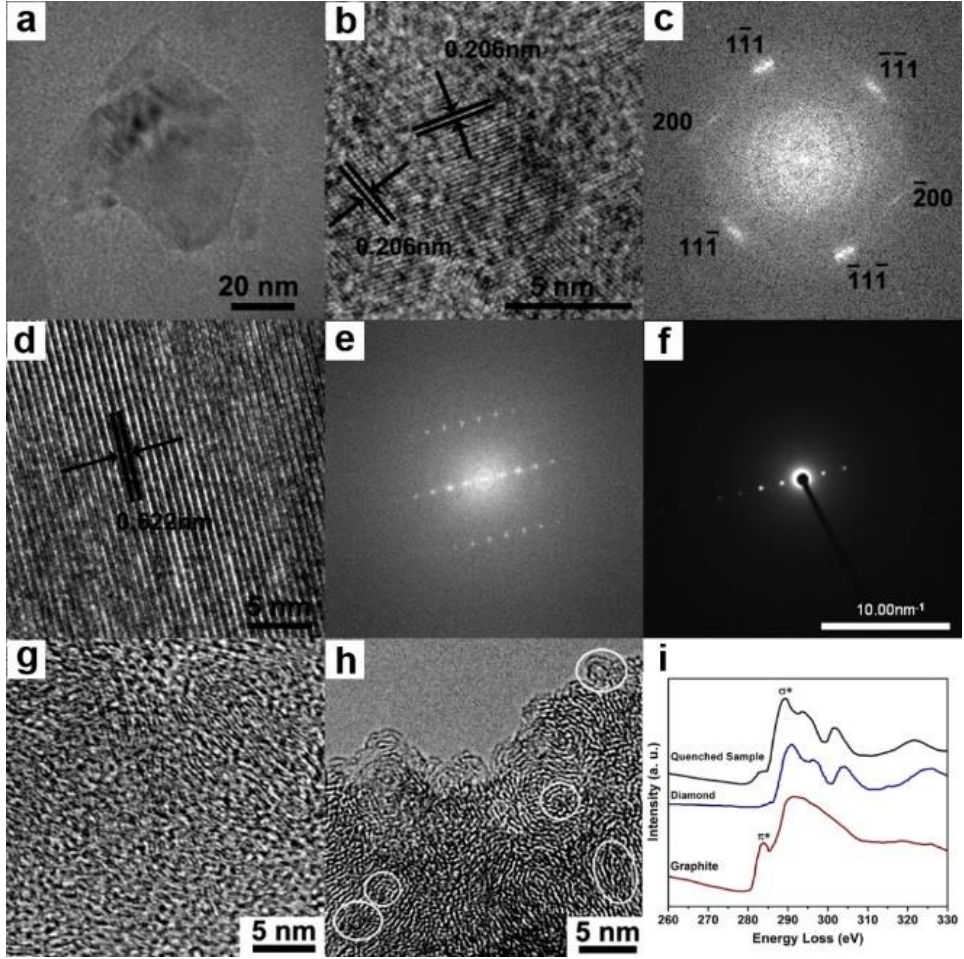

*Fig. 2: TEM analysis of the quenched sample. a. TEM image of a nanocrystalline cD, b. HRTEM image of the D nanocrystal, c. FFT in area showing crystal planes from image b, d. HRTEM of an orth-D. e. FFT of image d. f. SAED in the area of image d, g. HRTEM image of an amorphous phase, h. HRTEM image of fragmented G and fullerenes (circled). i. EELS in the area in image a and its comparison with D and G.*

Under low resolution transmission electron microscope (TEM) measurement of the quenched sample, we observed a nearly rectangular shaped crystal with edge lengths ~50nm (*Fig. 2a*). High resolution TEM (HRTEM) analysis of this crystal (*Fig. 2b, c*) showed a d-spacing of the displayed crystal planes of 2.06 Å with angles between the family planes estimated to be 68° to 70°, matching the d-spacing of the (111) planes and the calculated angles (70°) between the {111} family of a cD. Comparing the Electron Energy Loss Spectrum (EELS) of this crystal with pure nanocrystalline cD and G (Fig. 2i) shows that it contains mainly the spectrum of the σ* bond, similar to that of the nanocrystalline D with minimal π* bond character. The spectrum is identical to that in reference.[6] The π* bond may be from the residual G or other phases on or next to the D crystal. The crystal is evidently cD.

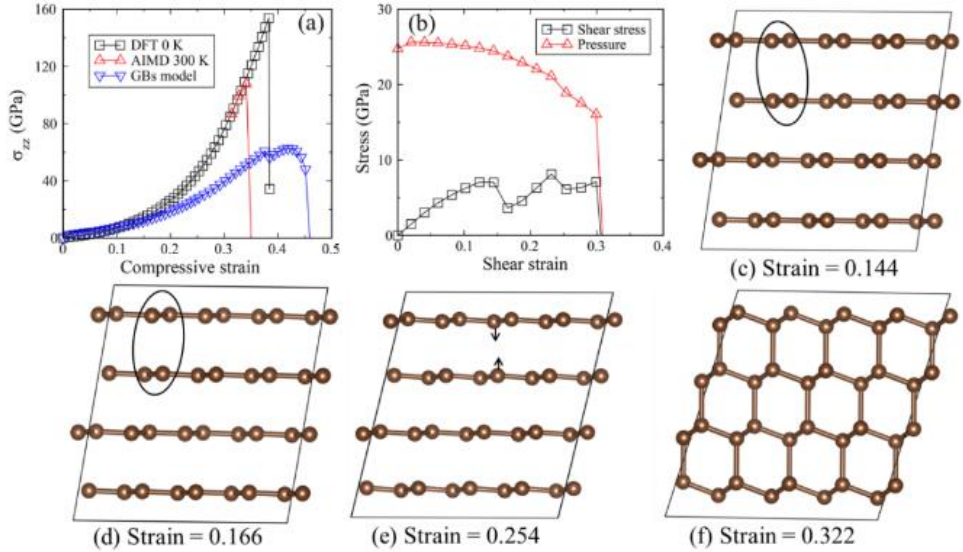

*Fig. 3: Stress-strain relationship and structural changes for compression and shear of G predicted from quantum mechanics and from grain boundary models for 30% pre-compressed G shearing along (0001)/<11$\bar{2}$0> slip system. (a) Stress-strain relationship for compression under various conditions; (b) stress-strain relationship for 30% pre-compressed G shearing along (0001)/<11$\bar{2}$0> slip system; (c) structure at shear strain 0.144; (d) structure at 0.166 shear strain showing slight shift of AB stacking layers (within the circles) along [12$\bar{3}$0] direction; (e) structure at shear strain 0.254 showing slight shift of C atoms along [0001] direction; (f) structure at 0.322 strain after transformation to hD phase. The arrows represent the C atom shifts along [0001] direction leading to the phase transition.*

The lowest transformation pressure from G to hD is 0.4 GPa and to cD is 0.7 GPa (Table S1 in Supplementary Materials (SM)). These are the lowest G to D transition pressures ever observed experimentally and are lowered by 50 and 100 times compared to those made under quasi-hydrostatic conditions, and much lower than ~20 GPa under plastic shear.[8, 9] Above all, the transition pressure is lower than the G-D phase equilibrium stress $\sigma_c = 2.239\ GPa$ with uniaxial compression and 1.939 GPa with superposed shear in addition (see SM).

In another region of the compressed-sheared G reaching 3 GPa, HRTEM study also indicates the formation of another phase (Fig. 2d, e, f). The d-spacings from selected area electron diffraction (SAED) and fast Fourier transformation (FFT) of the plane image (Table S2 in SM) can be indexed to an orthorhombic crystal structure, with cell parameters $a$ = 4.36, $b$ = 1.85, and $c$ = 12.50 Å. Most of the d-spacing4s are identical within the experimental error range to those in reference[10], especially those of (00$l$) planes. We therefore refer this phase to an ortho-D phase. Yet the diffraction pattern cannot fit into the proposed monoclinic structure (Fig. 2e), even though we believe this phase might be the same as the monoclinic D. The transition pressure to the ortho-D phase is also the lowest than those to form any D phases reported before.

Instability stress in density functional theory (DFT) study for pressure- and stress-induced transformations (SM) indicates superposed shear in a uniaxial compression of G could only reduce the transformation pressure from 250 GPa under hydrostatic conditions to about 20 GPa. In the

DFT simulation (SM) the G was initially subjected to 30% of uniaxial compression leading to uniaxial stress of 76.3 GPa. A shear along the (0001)/[11$\bar{2}$0] slip system causes no structural change before attaining 0.254 strain (67.2 GPa uniaxial stress) when the AB stacking layers are slightly shifted along the [11$\bar{2}$0] direction at 0.166 shear strain and 77.6 GPa uniaxial stress (Fig. 3(d)). At 0.254 strain, the carbon atoms within the plane are slightly shifted along the [0001] direction, forming out-of-plane deformation (Fig. 3(e)). At that point, application of an additional minimal shear increment of 0.045 (total strain of 0.299 and uniaxial stress of 63.5 GPa), the shifted carbon atoms are bonded to the neighbor layer, forming the D phase (Fig. 3(f)). In the calculation with a 20% of uniaxial pre-straining, no phase transition can be observed at all at shear strains to 0.82.

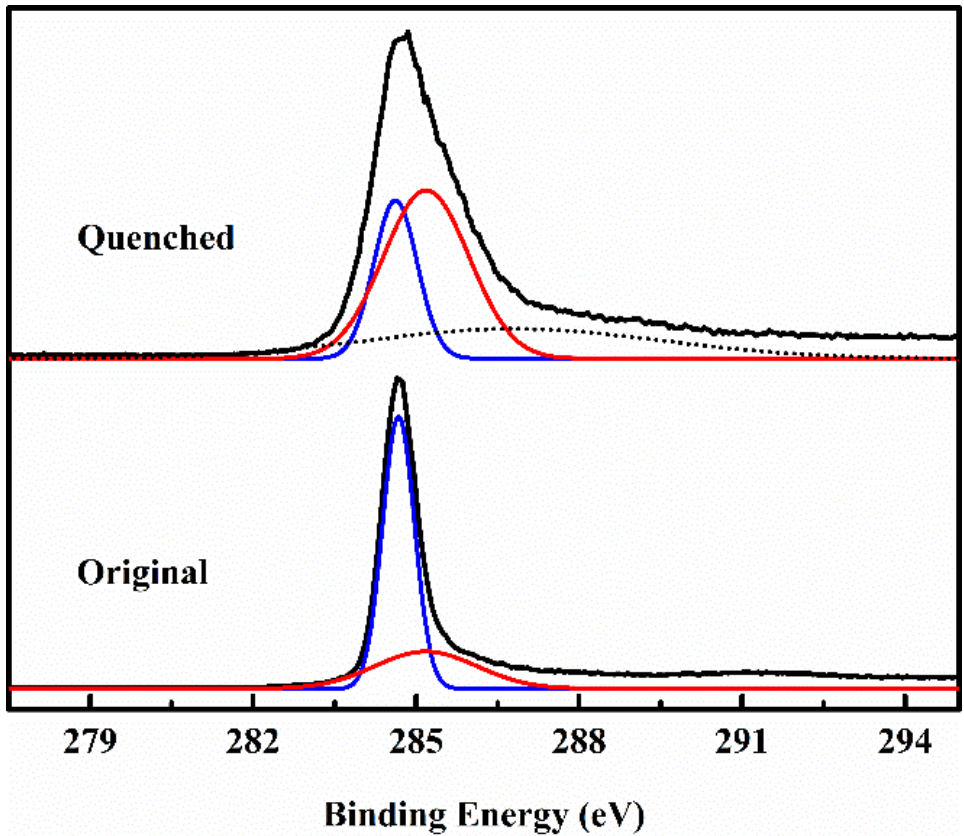

*Fig. 4: The XPS spectra of carbon of the sample before (bottom) and after (top) compression and shear. The blue, red and dashed lines represent the fitting results of $sp^2$, $sp^3$ and C-O XPS peak, respectively.*

There is a fundamental difference between pressure- and stress-induced transformations and strain-induced transformations under compression and plastic shear.[11, 12] Pressure- and stress-induced transformations occur by nucleation at pre-existing defects, e.g., dislocation pile-ups, which produce concentration of the stress tensor proportional to the number of dislocations. Strain-induced phase transformations occur by nucleation at new defects (dislocation pile-ups) generated during plastic deformation. Fragmentation of G into nanocrystals at the initial deformation stage provides obstacles for dislocation motion. Since number of dislocations in a strain-induced pile-up can be as large as 10 to 100, stresses at the tips can reach conditions for lattice instability (determined from the DFT simulations) even at small external pressure. The external pressure can be even smaller than the equilibirum pressure. Such a mechanism has been confirmed more quantitatively utilizing an analytical model of nucleation at the dislocation pile up in[11, 12] and phase field simulations of strain-induced transformations in a bi-crystal in[13, 14]. The G to D transition occurs without barrier as shear increases but nuclei cannot grow significantly because the stresses decrease with distance away from the defect tip. In addition, fragmentation form a large area of grain boundaries, where a high percentage of $sp^3$ bonding promotes out-of-plane deformations.

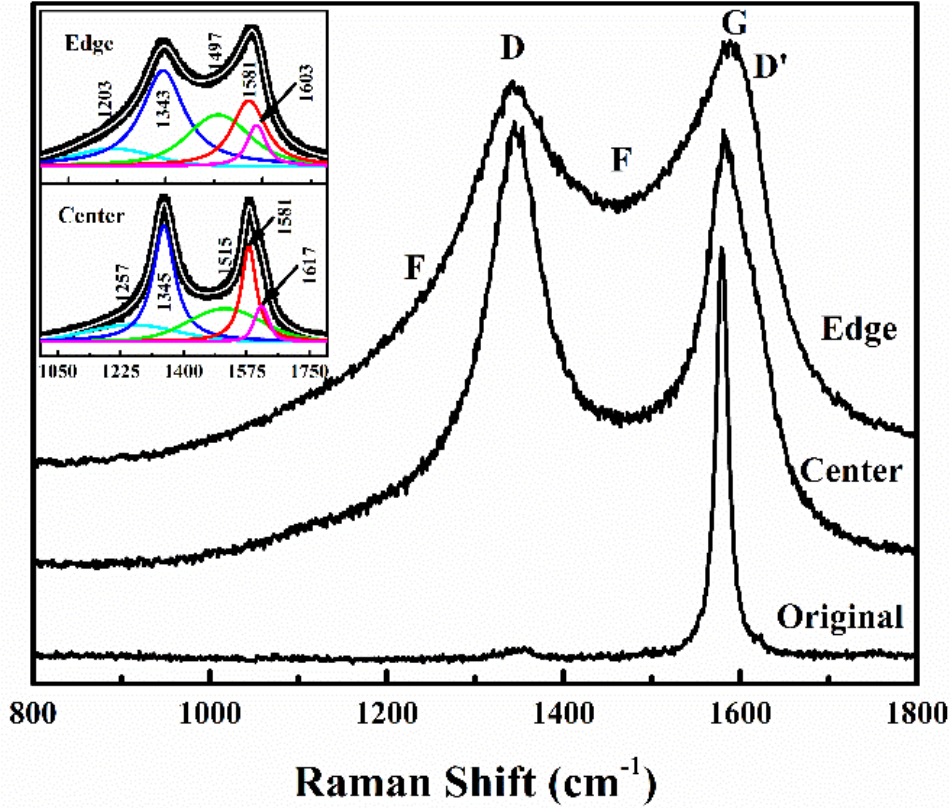

*Fig. 5: Raman spectra of original sample (bottom) and quenched sample at the center (middle) and edge (top) of the anvil. F, D, G, and D' respectively denote Fullerene, D, G, and D' bands. Inset, Voigt fitting of the Raman spectra of the quenched graphite. Black, white and colored lines are respectively the curves of measured spectra, the cumulative curve of fit results and the fit to Voigt peaks. The peaks at 1257 and 1203 cm$^{-1}$ (F band) and 1497 and 1515 cm$^{-1}$ are from the fullerenes; the peaks at 1345 and 1343 cm$^{-1}$ (D band), 1581 (G band), and 1603 and 1617 (D' band) are from nanocrystalline graphite; the peaks at 1497 and 1515 cm$^{-1}$ can also be ascribed to the formation of amorphous phases.*

*Formation of other phases.* From X-ray diffraction pattern of the quenched sample, we also observe a broad band ranging from 10 to 4 Å (Fig. 1b). HRTEM measurements show that in broader area the sample has turned into amorphous after compression/shear processes (Fig 2g). XPS measurements (Fig. 4) over large area of the quenched sample indicate that ratio of $sp^3$ to $sp^2$ bonding increased from 0.27 (in G) to 0.98 (Table S3 in SM). The amorphous phase has high $sp^3$ concentration and is thus believed to be diamond-like amorphous carbon. This is also confirmed by the Raman measurements (Fig. 5).

Near the edge of the sample in the HRTEM measurements, we also observed distorted and fragmented G planes, circular clusters, and fullerene-like structures. These structures randomly distributed in the sample without long range crystalline order. Thus, they also contribute to the broad band in the X-ray diffraction pattern. The Raman spectrum observed at the specific sites in the sample shows the spectrum of a fullerene as well.

The Raman spectroscopy of the sheet of the quenched sample shown in Fig. 5 exhibits broad F-, D-, G- and D'-bands.[15-18] Such bands are typically associated with the edge planes of G[17, 19] and the shortened cup-stacked type carbon nanotubes[20]. Along with the TEM observation, these bands are ascribed to the fullerenes, nanocrystalline G[19], and amorphous phases[21]. The applied shear in the experiment thus has transformed G into a variety of carbon types, all of which introduce much higher *sp³* concentration. Using the intensity of D- and G-bands, the cluster size of the G amorphous are estimated to be about 2 nm, which is also consistent with the observation in the TEM measurements.

Multiple phases coexisting in the reported experiment can be explained by different types of stress states at different defects and differently oriented grains, as well as by heterogeneity of the stress and plastic strain state in the sample, with maximum pressure at the center (which was measured) and maximum plastic strain at the edges, see Fig. S1 in SM. Such a broad spectrum of stress and plastic strain states allows us to obtain in a single experiment high throughput results with multiple phases from a single sample. Deviatoric (nonhydrostatic) stresses in a small region near tip of the defect are limited by the theoretical strength, which is one to two orders of magnitude larger than the macroscopic yield strength. Such a unique, highly deviatoric stress tensor at the limit of lattice instability is even not close to be approached otherwise. This creates unique opportunity for search for new phases near tips of strain-induced defects, which may not be obtained in bulk under hydrostatic conditions.

*Geological applications.* In geology, Ds are considered to be crystalized from various melts at temperature >900-1200$^0$C and pressure > 4 GPa in accordance with G-D equilibrium line, which corresponds to the depth of over 100 km of the Earth. Independent of this, micro Ds have been discovered in the low temperature and pressure continental crust and non-kimberlite bearing rocks, e.g., in the Kokchetav Massif of Russia,[22] the Dabie Shan mountains of China,[23] the Rhodope Mountains of Greece,[24] and the Variscan French Massif Central[25]. These micro Ds have long been interpreted as being of a metamorphic origin, that is the continental crust subducted to a depth of over 100 km and then uplifted to the surface by exhumation.[22-25] Our observation of D formation under shear at much lower pressure and 300 K demonstrates an alternative pathway for D formation in the interior of the Earth: carbon sources in the crust of the Earth may transform directly to D under sufficient shear strain. Thus, D can be formed in the crust under geological processes involving considerable relative motion activities. For instance, the D in the Kokchetav Massif may have formed within shear bands during tectonic rifting in the Mesozoic Era. The D in the Dabie mountain may have formed in a similar historical event during the collision of the Yangtze plate against the North China plate. Such shear-induced transformations may resolve the riddle that micro D sites lack a sedimentary lithosphere. Even with subduction, micro D can be formed at much shallower depth since shear is generated during the subduction process. Also, knowing that shear strain can induce formation of diamond at low pressure and temperature suggests new strategies for seeking diamond in the crust.

*Concluding Remarks.* Applying plastic shear, we synthesized hD and cD at pressures as low as 0.4 to 0.7 GPa and 300 K, two orders of magnitude below transformation pressures under quasi-hydrostatic conditions and even below the phase equilibrium. We show also that shear leads to other Ds and fullerene phases at modestly higher pressure. We use theory to explain that this drastically reduced pressure for transformation from crystalline G to crystalline D arises from the strong stress tensor concentration (at the limit of lattice instability) at the tip of plastic strain-induced defects. Thus, we provide and validate experimentally a new mechanism for phase transformations and for materials synthesis. Thus, instead of the pressure of 5 GPa, temperature of 1500 K, plus catalyst, we show that applying a pressure of just 0.7 GPa at room temperature in the presence of large plastic shear deformation (by one of the traditional methods, like high-pressure torsion, extrusion, or ball milling) is sufficient. Moreover, our findings open many fundamental questions about multiscale effect of nonhydrostatic stresses, plastic strains, and defects on high-pressure phase transformations. The effect of shear on reducing the transition pressure for G to D suggests re-investigation of the role of pressure versus shear for phenomena at the interior of the Earth and other planets. In particular, the geological mechanisms related to micro D formation in the crust under low pressure and temperature can be revised.

**Acknowledgement**

This work was supported by National Science Foundation (Grant No. DMR1431570, program manager, John Schlueter). V.I.L. and B.F. also acknowledge support from Army Research Office (Grant No. W911NF-17-1-0225 managed by David Stepp) and Vance Coffman Faculty Chair Professorship. Synchrotron X-ray experiment was performed at Cornell High Energy Synchrotron Source. The authors thank Dr. Zhongwu Wang for experimental technical support.

**Supplementary Materials Reference** [11, 12, 26-32]

# Supplementary Materials

### Shear experiments details and Diamond Observation

The compression and shear experiments were performed using a rotational apparatus with two anvils made of polycrystalline cubic boron nitride that are oppositely aligned with backing load for pressure generation and relative rotation of one to generate shear on powder sample in between. The G powder sample was loaded between anvil flat tops of 3 mm diameter. The sample thickness after loading was ~50 μm, and a small piece of gold particle with dimensions of 20 μm, was placed on top of the sample at the center for pressure calibration. Synchrotron X-ray diffraction was performed after each operation. Based on the volume change of gold after each processing, the pressure is determined by its equation of state.[26]

**Table S1: The phases identified by XRD with sequential compression and shear operation**

| Run 1 | | | | Run 2 | | | | Run 3 | | | |
|---|---|---|---|---|---|---|---|---|---|---|---|
| L ($10^2$kg) | R (°) | P (GPa) | Phase | L ($10^2$kg) | R (°) | P (GPa) | Phase | L ($10^2$kg) | R (°) | P (GPa) | Phase |
| 0 | 0 | 0 | No | 0 | 0 | 0 | No | 6.5 | 0 | 0 | No |
| 6.5 | 0 | 0.1 | No | 14 | 45 | 1.2 | H | 6.5 | **45** | **0.7** | **C** |
| 8 | **45** | **0.4** | **H** | 12 | 45 | 1.2 | H | 6.5 | 135 | 1.0 | C |
| 13 | 45 | 0.9 | C | 12 | 180 | 3.1 | H | 6.5 | 225 | 0.3 | No |
| 14 | 45 | 4.4 | C | 12 | 195 | 3.1 | No | 6.5 | 315 | 2.3 | No |
| 14 | 180 | 3.2 | C | 12 | 210 | 3.0 | C | 6.5 | 395 | 2.3 | C |
| 14 | 315 | 8.0 | H,C | 12 | 225 | 3 | C | 6.5 | 440 | 2.4 | C |
| 14 | 450 | 2.3 | H,C | 12 | 270 | 2.8 | C | 6.5 | 530 | 3.0 | H |
| 14 | 630 | 0.6 | H | 12 | 315 | 3.2 | C | 6.5 | 665 | 2.4 | H |
| 14 | 630 | 1.7 | H | 12 | 405 | 2.9 | No | 6.5 | 775 | 2.4 | H,C |
| 14 | 990 | 1.0 | H | 12 | 440 | 2.9 | No | 6.5 | 845 | 2.5 | C |
| 14 | 1170 | 1.8 | H | 10 | 450 | 2.9 | C | 6.5 | 1025 | 2.2 | C |
| 0* | 0 | 0 | C | 10 | 620 | 3.3 | C | 6.5 | 1070 | 2.4 | No |
| | | | | 10 | 710 | 3.3 | C | 6.5 | 1072 | 2.4 | No |
| | | | | 10 | 845 | 2.8 | C | 0* | 0 | 0 | C |
| | | | | 10 | 1205 | 2.5 | C | | | | |
| | | | | 10 | 1565 | 2.0 | C | | | | |
| | | | | 10 | 1925 | 2.0 | No | | | | |
| | | | | 0* | 0 | 0 | C | | | | |

*, From the recovered quenched sample (without anvil in XRD measurements).

H and C represent hexagonal and cubic diamonds, respectively.

Run I, measurements with diamond composite anvils; Run II and III, measurements with cubic boron nitride anvils. L is the axial load, R is an anvil rotation, P is the pressure determined though the equation of state of gold with unit cell volume determined from observed X-ray diffraction, Phase, the phase other than G observed by X-ray, C for cD and H for hD; No, H and C were not observed by X-ray beam.

From the listed operation conditions (Table S1), we found that the lowest pressure for the formation of hD was 0.4 GPa, and 0.7 GPa for cD. Note that "No" in Table S1 under pressure and shear after D were detected at the previous operations may be due to that the D were moved away from the x-ray beam. We observed cD from the quenched sample along with amorphous phases, but did not observe any diffraction lines for hD. The transformation to hD is reversible after treatment at even larger pressure. It might also transform to cD after additional processing.

## 1. *In-situ* X-ray diffraction experiments

The in-situ X-ray diffraction experiment was performed at B1 station in Cornell High Energy Synchrotron Source (CHESS) with X-ray wavelength of 0.4859 Å.

## 2. Raman spectroscopy

The Raman spectroscopy was performed using a Renishaw InVia Raman spectrometer and laser wavelength of 532 nm.

## 3. X-ray photoelectron spectroscopy

The XPS was performed using a Thermo Scientific ESCALAB 250 High Performance Imaging XPS apparatus using monochromatic Al Kα X-ray source ($h\upsilon=1486.6\ eV$). All XPS data were acquired at a nominal photoelectron take off angle of 55˚.

## 4. High-resolution transmission electron microscope

The high-resolution transmission electron microscope was a Hitachi H-8100 IV TEM operating at 200 kV accelerating voltage.

**Table S2: Index of the orthorhombic phase in comparison with the reported monoclinic phase.**

| Observed d-spacings (nm) | | Miller index (hkl) | |
|---|---|---|---|
| This study | Reference * | Monoclinic** | Orthorhombic (this study) |
| 0.622 | 0.6239 | 0 0 2 | 0 0 2 |
| 0.311 | 0.3120 | 0 0 4 | 0 0 4 |
|  | 0.2179 | 2 0 0 | 2 0 0 |
|  | 0.2142 | 1 $\bar{1}$ 1 | 1 0 5 |
| 0.208 | 0.2080 | 0 0 6 | 0 0 6 |
|  | 0.2068 | 2 0 $\bar{2}$ | 2 0 $\bar{2}$ |
|  | 0.2048 | 2 0 2 | 2 0 2 |
| 0.183 |  |  | 0 1 $\bar{1}$ |
| 0.183 |  |  | 0 1 1 |
|  | 0.1800 | 2 0 $\bar{4}$ | 2 0 $\bar{4}$ |
|  | 0.1774 | 2 0 4 | 2 0 4 |
| 0.169 |  |  | 0 1 $\bar{3}$ |
| 0.168 | | | 0 1 3 |
| | 0.1647 | 1 1 5̄ | 0 1 7 |
| 0.156 | | | 0 0 8 |
| | 0.1517 | 2 0 6̄ | 2 0 6̄ |
| 0.150 | | | 0 1 5 |
| | 0.1493 | 2 0 6 | 0 2 9 |
| 0.148 | | | 0 1 5̄ |
| | 0.1278 | 2 0 8̄ | 2 0 8̄ |
| | 0.1259 | 2 0 8 | 2 0 8 |
| | 0.1258 | 3 1̄ 0 | 3 0 5 |
| 0.124 | | | 0 0 10 |
| | 0.1090 | | 0 1 9 or 2 1 7 |

\*, data from Reference.(10)

\*\*, Monoclinic unit cell (reference): $a = 0.436$ nm, $b = 0.251$ nm, $c = 1.248$ nm, $\beta = 90.9°$.(10)

\*\*\*, Orthorhombic unit cell (this study): $a = 0.436$ nm, $b = 0.185$ nm, $c = 1.25$ nm.

**Table S3: X-ray Photon spectrum analysis of the sample before and after shear processing.**

| Sample | Bonding type | Bonding energy (eV) | Intensity | Ratio to sp² |
|---|---|---|---|---|
| Starting graphite | C-$sp^2$ | 284.68 | 165394 | 1 |
| | C-$sp^3$ | 285.26 | 43828 | 0.27 |
| Quenched sample | C-$sp^2$ | 284.68 | 59208 | 1 |
| | C-$sp^3$ | 285.26 | 59215 | 0.98 |
| | C-O | 286.50 | 34293 | 0.58 |

**Determination of equilibrium stress under nonhydrostatic conditions**

The phase equilibrium pressure of G and D is $p_e = 1.7$ GPa at 0 K and 2.45 GPa at 300 K.(8) Yet the G to D solid-solid phase transformation pressures observed in the past are more than an order of magnitude higher due to the large energy barrier. Strong effect of nonhydrostatic loading is expected due to very anisotropic (i.e., with large deviatoric part) lattice deformation during G to D transformation. The components of the transformation deformation gradients are $F_c = 0.627$ in the $c$ direction and $F_l = 1.025$ in the two-lateral direction, i.e., volumetric transformations gradient is $F_V = F_c F_l^2 = 0.659$. Thus, lateral compression suppresses transformation. Equaling transformation work under hydrostatic compression by pressure $p$ and uniaxial stress along $c$ direction $\sigma_c$, $\sigma_c(1 - F_c) = p_e(1 - F_V)$, we obtain $\sigma_c = 0.914 p_e$ and corresponding pressure $p_c = \sigma_c/3 = 0.305 p_e$. Consequently, if in experiments the equilibrium pressure is recorded, then it reduces by a factor of more than 3. However, still one has to apply an axial stress (force), and it is almost the same ($\sigma_c = 0.914 p_e$) as for hydrostatic conditions. Here, we neglected the contribution of jump in elastic moduli to the mechanical driving force for a transformation in comparison with the transformation work, as well as small increase in the area orthogonal to *c* axis, which does not change conclusion. The transformation shear for transformation from G to cubic D is ~0.3. Superposing shear stress $\tau$ reduces $\sigma_c$ to $\sigma_c = 0.914 p_e - 0.3\tau$ and $p_c = 0.305 p_e - 0.1\tau$. Taking equilibrium $p_e = 2.45\ GPa$ and $\tau = 1\ GPa$ (limited by the macroscopic yield strength in shear of G), we obtain $p_c = 0.747\ GPa$ ($\sigma_c = 2.24\ GPa$) for uniaxial compression and $p_c = 0.647\ GPa$ ($\sigma_c = 1.94\ GPa$) under compression and shear. Thus, despite the overestimated yield strength in shear of the G, the effect of shear stress on equilibrium pressure is small. In general, even for such large deviatoric transformation strain, the effect of nonhydrostatic loading on the measurable equilibrium axial stress $\sigma_c$ (2.45 vs. 1.94 GPa) is relatively modest and cannot explain drastic reduction in transformation pressure. Also, in experiment, there are the lateral stresses that increase the equilibrium stress and pressure; making it more reasonable to compare experimental values with $\sigma_c$ rather than $p_c$. Important point is that both hD and cD appeared in experiment at pressure well below the equilibrium stress, even with allowing for the effect of nonhydrostatic stresses.

**Atomistic Simulations**

All simulations were performed using the Vienna Ab-initio Simulation Package (VASP) periodic code.[27] Plane wave basis sets were chosen to expand the Kohn-Sham eigenfunctions. VASP used a projector-augmented-wave approach (PAW) for describing the electron-ion interaction. The Perdew–Burke–Ernzerhof (PBE) functional is used accounting for the exchange-correlation of electron-electron interactions. The London dispersion (van der Waals attraction) is treated by Grimme DFT-D3 approach. An energy cutoff of 500 eV is used in all the simulations since it gives excellent convergence on energy, force, stress, and geometries. The energy error for terminating electronic self-consistent field (SCF) and the force criterion for the geometry optimization were set equal to $10^{-6}$ eV and $10^{-3}$ eV/Å, respectively. Reciprocal space was sampled using the Γ-centered Monkhorst−Pack scheme with a fine resolution of $2\pi \times 1/40$ Å$^{-1}$ for all calculations except for the large grain boundary model with 152 atoms. The Monkhorst-Pack grid (2×2×1) in the k-space was used for the GB model.

**Instability stresses for pressure- and stress-induced transformations**

The density functional theory (DFT) method was used to systematically determine the instability stress (stress discontinuity) to reflect stress-induced phase transition of G. The instability pressure for perfect G crystal under hydrostatic condition is $p_i = 250\ GPa$ (at 0 K), which greatly exceeds $p_e$. For uniaxial compression along the *c*-axis ([0001] direction) of G (with zero lateral strains), $p_i = 52\ GPa$. Addition of a shear stress in the range 6-8 GPa under periodic conditions at lateral sides makes $p_i = 17 - 26\ GPa$. The DFT-MD simulations indicate that temperature increment of 300 K leads to $p_i = 15\ GPa$, due to thermal fluctuations even though it increases $p_e$. A pentagon/heptagon type grain boundary in a G crystalline reduces transformation pressure from 51 to 20 GPa in a uniaxial compression. Even though this clearly suggests that elastic shear deformation can dramatically decrease the transition (instability) pressure, the pressure in the experiment is substantially lower than the calculated transition pressure because of lack of stronger defects, like dislocation pile-ups. In the case where initial 30% of uniaxial compression applied to a single crystal graphite, a shear of 0.254 along the (0001)/[11$\bar{2}$0] slip system causes the carbon atoms within the plane being slightly shifted along the [0001] direction, forming out-of-plane deformation. At that point, application of an additional minimal shear increment of 0.045, the shifted carbon atoms are bonded to the neighbor layer, forming the D phase.

**Molecular dynamics simulation**

To examine the structural changes at finite temperature, we carried out Molecular Dynamics using Quantum mechanics forces (AIMD). In each compressive strain step, the systems were equilibrated at 300 K for 1 ps using the NVT (constant volume, constant temperature, and constant number of atoms) ensemble. This leads to a strain rate of $2.0 \times 10^{10}$ s$^{-1}$ for compressive deformation. The time constant for the Nose thermostat was 0.1 ps and the time step 1 fs was used for integrating the equation of motion. We used the last 0.5 picoseconds to compute the stress statistically in each strain step.

**Strain-induced transformations to and from amorphous phases**

Large plastic straining can cause amorphization of G, similar to amorphization in many other materials. Amorphous D may appear through strain-induced transformation from crystalline G at the tip of the dislocation pile-up or other strong defects, or, at pressure exceeding the equilibrium pressure for amorphous phases, during plastic deformation of amorphous G. In turn, both amorphous G and D may undergo a reconstructive transformation under large shear into more stable cubic D. The main mechanism of plastic deformation in amorphous materials is related to a series of irreversible atomic rearrangements within small localized shear transformation zones.[28-30] These atomic rearrangements play the role similar to the thermal fluctuations at high temperature toward a stable or metastable phases, thus overcoming energy barrier.

**Micro- and macroscale modeling of strain-induced phase transformations between G and cD and hD**

A micro- and macroscale modeling provides further insight in the possible heterogeneous kinetics of phase transformations, which is conceptually different from the traditional kinetics under quasi-hydrostatic conditions. Strain-induced transformations should be characterized in terms of strain controlled kinetic equations for concentrations of all participating phases. We will focus on two transformations: G→hD and G→cD; the second transformation includes G→hD →cD, which is why we will not consider transformation hD →cD separately. General kinetic equations are presented in [11, 12, 28]). They are based on the coarse graining of our nanoscale model of nucleation on strain-induced defects.[11, 12] In particular, kinetics is formulated in terms of derivative of the concentration of phases with respect to accumulated plastic strain $q$ (rather than time) and it depends on pressure $p$ and on the ratio of the yield strengths of all phases, because plastic strain localizes in the phases with the smaller yield strength. Phase equilibrium and lattice instability pressures are not present in these equations. Instead, the minimum pressures for different strain-induced transformations, which are calibrated from experiment and take into account the strongest defects responsible for nucleation, participate in the model. The microscale kinetic equations are included in a macroscale model for coupled plastic flow and strain-induced phase transformation.[31, 32]

Finite element solution for compression and shear of a sample in RDAC is presented in Fig. S1. Plastic strain is distributed relatively homogeneously after compression. After torsion, it grows with increasing radius within transformed region, then oscillates. This is surprising for shearing, which is proportional to the radius. The reason is in large reduction in thickness during rotation of an anvil (from 300 to 200 µm) and heterogeneous relative sliding of material with respect to an anvil. Important point is that the pressure is maximum at the center, where it is measured, i.e., reported transformation pressures are the upper bounds over all possible ones in the sample. Distribution of concentration of D phases are determined mostly by pressure distribution, i.e., they increase toward the sample center. Pressure drops in the major part of a sample during rotation of an anvil due to large volume reduction during phase transformations. This does not contradict to the constant axial force, because force is determined by the axial stresses rather than pressure. Simulation results are in qualitative agreement with experiments, namely, that Ds do not appear under compression but appear under torsion (while pressure is getting smaller), hD appears earlier but then the concentration of cD is getting larger, and that pressure at the center does not change significantly during rotation by 45º.

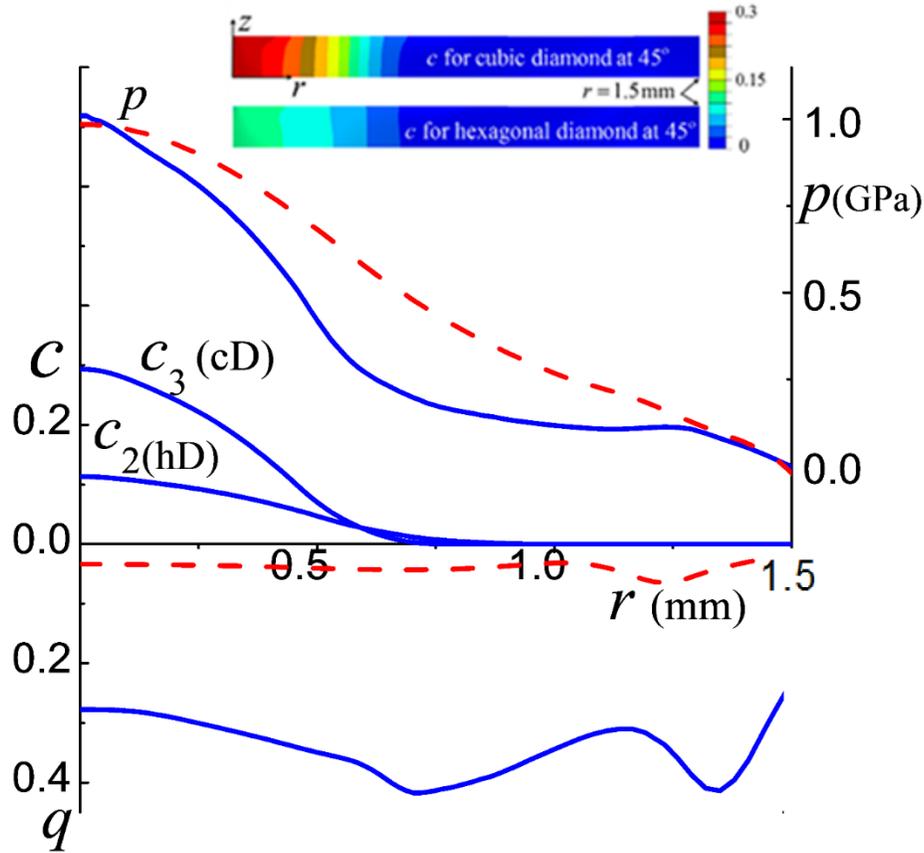

*Fig. S1: Distributions of the fields along the sample radius under compression and torsion in RDAC. Distributions of the accumulated plastic strain (q), concentrations $c_2$ and $c_3$ of the hD and cD, respectively, and pressure p. The dashed lines are after compression and solid lines are for torsion by 45º at constant force. Inset shows distribution of concentrations $c_2$ and $c_3$ in the cross section of a sample, which vary weakly along the sample thickness.*